\documentclass[]{pasj01_letter}

\usepackage{color}
\usepackage[normalem]{ulem}
\usepackage[]{graphicx}
\newcommand{\red}{\textcolor{black}}
\usepackage{multirow}

\begin{document} 

\title{ALMA CO Observations of a Giant Molecular Cloud in M33: Evidence for High-Mass Star Formation Triggered by Cloud-Cloud Collisions}

\author{Hidetoshi \textsc{SANO}\altaffilmark{1}}%
\author{Kisetsu \textsc{TSUGE}\altaffilmark{2}}%
\author{Kazuki \textsc{TOKUDA}\altaffilmark{1,3}}%
\author{Kazuyuki \textsc{MURAOKA}\altaffilmark{3}}%
\author{Kengo \textsc{TACHIHARA}\altaffilmark{2}}%
\author{Yumiko \textsc{YAMANE}\altaffilmark{2}}%
\author{Mikito \textsc{KOHNO}\altaffilmark{2,4}}%
\author{Shinji \textsc{FUJITA}\altaffilmark{2}}%
\author{Rei \textsc{ENOKIYA}\altaffilmark{2}}%
\author{Gavin \textsc{ROWELL}\altaffilmark{5}}%
\author{Nigel \textsc{MAXTED}\altaffilmark{6}}%
\author{Miroslav D. \textsc{FILIPOVI{\'C}}\altaffilmark{7}}%
\author{Jonathan \textsc{KNIES}\altaffilmark{8}}%
\author{Manami \textsc{SASAKI}\altaffilmark{8}}%
\author{Toshikazu \textsc{ONISHI}\altaffilmark{3}}%
\author{Paul P. \textsc{PLUCINSKY}\altaffilmark{9}}%
\author{Yasuo \textsc{FUKUI}\altaffilmark{2,10}}%
\email{hidetoshi.sano@nao.ac.jp}
\altaffiltext{1}{National Astronomical Observatory of Japan, Mitaka, Tokyo 181-8588, Japan}
\altaffiltext{2}{Department of Physics, Nagoya University, Furo-cho, Chikusa-ku, Nagoya 464-8601, Japan}
\altaffiltext{3}{Department of Physical Science, Graduate School of Science, Osaka Prefecture University, 1-1 Gakuen-cho, Naka-ku, Sakai, Osaka 599-8531, Japan}
\altaffiltext{4}{Astronomy Section, Nagoya City Science Museum, 2-17-1 Sakae, Naka-ku, Nagoya, Aichi 460-0008, Japan}
\altaffiltext{5}{School of Physical Sciences, The University of Adelaide, North Terrace, Adelaide, SA 5005, Australia}
\altaffiltext{6}{School of Science, University of New South Wales, Australian Defence Force Academy, Canberra, ACT 2600, Australia}
\altaffiltext{7}{Western Sydney University, Locked Bag 1797, Penrith South DC, NSW 2751, Australia}
\altaffiltext{8}{Dr. Karl Remeis-Sternwarte, Erlangen Centre for Astroparticle Physics, Friedrich-Alexander- Universit$\ddot{a}$t Erlangen-N$\ddot{u}$rnberg, Sternwartstra$\beta$e 7, D-96049 Bamberg, Germany}
\altaffiltext{9}{Harvard-Smithsonian Center for Astrophysics, 60 Garden St., Cambridge, MA 02138, USA}
\altaffiltext{10}{Institute for Advanced Research, Nagoya University, Furo-cho, Chikusa-ku, Nagoya 464-8601, Japan}

\KeyWords{ISM: H{\sc ii} regions---Stars: formation---ISM: individual objects (M33, M33GMC~37)}

\maketitle

\begin{abstract}
We report the first evidence for {high-mass} star formation triggered by collisions of molecular clouds in M33. Using the Atacama Large Millimeter/submillimeter Array, we spatially resolved filamentary structures of giant molecular cloud 37 in M33 using $^{12}$CO($J$ = 2--1), $^{13}$CO($J$ = 2--1), and C$^{18}$O($J$ = 2--1) line emission at a spatial resolution of $\sim2$ pc. There are two individual molecular clouds with a systematic velocity difference of $\sim6$ km s$^{-1}$. {T}hree continuum sources {representing up to $\sim10$ {high-mass} stars with} the spectral types of {B0V--O7.5V} are embedded within the densest parts of molecular clouds bright in the C$^{18}$O($J$ = 2--1) line emission. The two molecular clouds show a complementary spatial distribution with a spatial displacement of {$\sim6.2$} pc, and show a V-shaped structure in the position-velocity diagram. These observational features {traced by CO and its isotopes} are consistent with {those in} {high-mass} star-{forming regions created by cloud-cloud collisions in} the Galactic and Magellanic {Cloud} H{\sc ii} regions. {Our new finding in M33 indicates that the cloud-cloud collision is a promising process to trigger {high-mass} star formation in the Local Group.}
\end{abstract}

\section{Introduction}
It is a long-standing question how {high-mass} stars are formed in galaxies. Three different {hypothesis} are thought to be mechanisms of the {high-mass} star formation: the monolithic collapse, competitive accretion, and the stellar mergers (e.g., \cite{2007ARA&A..45..481Z} and references therein). However, these theoretical models have remained controversial due to lacking conclusive observational evidence. 

Recently, an alternative idea of {high-mass} star formation triggered by ``cloud-cloud collisions'' has received much attention since the discovery of 50 or more pieces of observational evidence {(e.g., \cite{2018ApJ...859..166F,2019PASJ..tmp..127E} and references therein)}. The colliding clouds have a supersonic velocity {difference} with an intermediate velocity component---bridging feature---{created} by the collisional deceleration. The complementary spatial distribution of two clouds {(or spatial anti-correlation between the two clouds)} is one of the important signatures of collisions because one of the colliding clouds can create a hollowed-out structure in the other cloud. Furthermore, a V-shaped structure can be seen in the position-velocity diagram due to the deceleration and hollowed-out structure. Theoretical studies also support these observational signatures, and {predict that} cloud-cloud collision increases the {effective Jeans mass} {so high-mass stars can} form {the high-mass stars} in the shock-compressed layer (e.g., \cite{1992PASJ...44..203H,2010MNRAS.405.1431A,2013ApJ...774L..31I,2014ApJ...792...63T,2018PASJ...70S..54S,2018PASJ...70S..53I}).

Investigating {the} universality of the cloud-cloud collision scenario, we need further observational examples under various environments and their scales. {In t}he individual {G}alactic star-forming regions such as Orion and Vela, the {high-mass} star formation can be understood by at least five cloud-cloud collisions {on} $\sim$1--100 pc scales (\cite{2016ApJ...820...26F}, \yearcite{2018ApJ...859..166F}; \cite{2017arXiv170605664F,2018PASJ...70S..43S,2018PASJ...70S..48H,2018PASJ...70S..49E}). It is remarkable that collisions of molecular clouds are seen not only in our Milky Way, but also in the {Magellanic Clouds} ({e.g.,} \cite{2015ApJ...807L...4F}; \cite{2017ApJ...835..108S}). Furthermore, the tidally driven galactic-scale ({a few kpc scales}) collisions of H{\sc i} clouds are found in the Magellanic and M31--M33 systems (\cite{2017PASJ...69L...5F}, {\yearcite{2019ApJ...886...14F};} \cite{2019ApJ...871...44T,2018PASJ...70S..52T}{; \cite{2019ApJ...886...15T}}). We can therefore reveal sites of cloud-cloud collisions even in external galaxies if the spatial resolution of CO/H{\sc i} data is high enough {(e.g., a few pc scales)}.

{Here, we} report the first evidence for the {high-mass} star formation triggered by collisions of molecular clouds in M33. {Sections \ref{obs} and \ref{results} describe observations and data reduction of the ALMA CO and continuum datasets and their results. Section \ref{spectral} gives properties of {high-mass} stars {in the region}; Section \ref{ccc} presents a possible scenario of cloud-cloud collision as the formation mechanism of {high-mass} stars. A summary and conclusions are provided in Section \ref{conclusion}.}

\begin{figure*}
\begin{center}
\includegraphics[width=\linewidth]{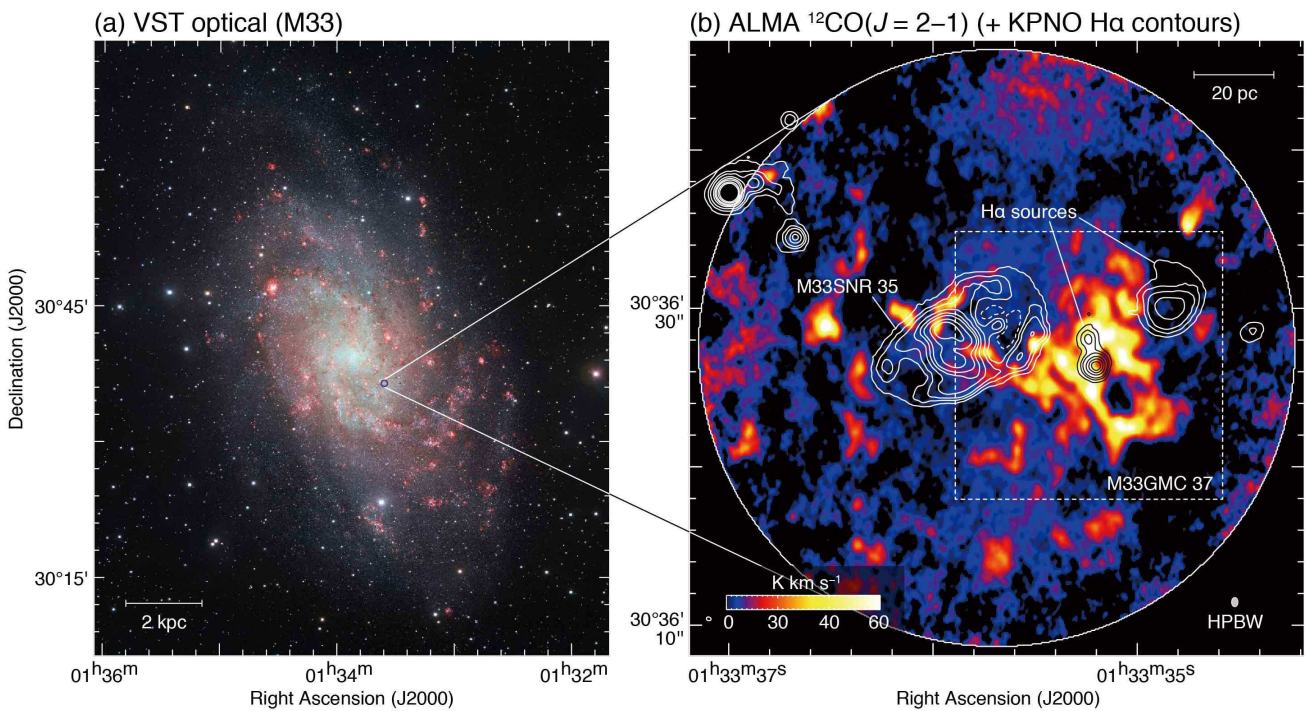}
\end{center}
\caption{{False color image of M33 obtained with the VLT Survey Telescope (VST, credit: ESO). The red, orange, and cyan represent H$\alpha$, r-band , and g-band, respectively. The blue circle indicates the ALMA FoV. The scale bar is shown in the bottom left corner.} (b) Integrated intensity map of $^{12}$CO($J$ = 2--1) obtained with ALMA. The integration velocity range is $-145$--$-126$ km s$^{-1}$. The superposed {white and black} contours indicate the H$\alpha$ intensity obtained by the Kitt Peak National Observatory (KPNO, \cite{2006AJ....131.2478M}). The contour levels are 300, 400, 500, 700, 900, 1100, 1500, and 1900 ergs s$^{-2}$ cm$^{-2}$. The dashed box containing M33GMC~37 represents the region to be shown in Figures \ref{fig2}, \ref{fig3}, and \ref{fig6}. We also annotated positions of M33SNR~35 and H$\alpha$ sources. {The scale bar and beam size are also shown in the top right and bottom right corners, respectively.}}
\label{fig1}
\end{figure*}%

\section{Observations and Data Reduction}\label{obs}
We carried out ALMA Band~6 {(211--275~GHz)} observations toward M33GMC~37 in Cycle~6 as part of the CO survey toward the SNRs in M33 (PI: H. Sano, \#2018.1.00378.S). We used the single-pointing observation mode with 45--49 antennas of the 12-m arrays. The center position of the pointing is ($\alpha_\mathrm{J2000}$, $\delta_\mathrm{J2000}$) $\sim$ ($01^\mathrm{h}33^\mathrm{m}35 \fs9$, $30\degree36\arcmin27\farcs5$). {The baseline length ranges from 15.06 to 1397.85 m, corresponding to {\it{u--v}} distances from 11.6 to 1074.9 $k \lambda$. The maximum recoverable scale (a.k.a. largest angular scale) is calculated to be 3\farcs73. Two quasars, J2253$+$1608 and J0237$+$2848 were observed as bandpass and flux calibrators. We also observed the quasar J0112$+$3208 as a phase calibrator.} There were two spectral windows including the $^{12}$CO($J$ = 2--1), $^{13}$CO($J$ = 2--1), and C$^{18}$O($J$ = 2--1) line emission with a bandwidth of 117.19~MHz. The frequency resolution was 70.6~kHz for $^{12}$CO($J$ = 2--1) and 141.1~kHz for $^{13}$CO($J$ = 2--1) and C$^{18}$O($J$ = 2--1). We also observed two spectral windows as continuum bands, of which frequency ranges are 231.0--233.0~GHz and 216.3--218.2~GHz. {Although these continuum bands contain line emission {of} H(30${\alpha}$) and SiO($J$ = 5--4), we could not detect the two lines significantly.}

The data reduction was performed using the Common Astronomy Software Application (CASA; \cite{2007ASPC..376..127M}) package version 5.5.0. We used the ``multiscale CLEAN'' algorithm implemented in the CASA package (\cite{2008ISTSP...2..793C}). The synthesized beam of final dataset is $0\farcs59 \times 0\farcs42$ with a position angle (P.A.) of $0\fdg4$ for $^{12}$CO($J$ = 2--1), $0\farcs62 \times 0\farcs44$ with a P.A. of $0\fdg1$ for $^{13}$CO($J$ = 2--1), $0\farcs63 \times 0\farcs44$ with a P.A. of $1\fdg1$ for C$^{18}$O($J$ = 2--1), and $0\farcs59 \times 0\farcs43$ with a P.A. of $-1\fdg4$ for the 1.3 mm continuum. The typical spatial resolution is $\sim2$~pc at the distance of M33 {($817 \pm 59$ kpc, \cite{2001ApJ...553...47F})}. The typical noise fluctuations of the line emission and continuum are $\sim0.15$~K at the velocity resolution of 1 km s$^{-1}$ and 0.017 mJy beam$^{-1}$, respectively. To estimate the missing flux {density}, we used the $^{12}$CO($J$ = 2--1) dataset obtained with the IRAM 30-m radio telescope (\cite{2010A&A...522A...3G,2014A&A...567A.118D}). {Following the methods of \citet{2014A&A...567A.118D}, we applied a forward efficiency of 0.92 and a main beam efficiency of 0.56 to convert the main beam temperature scale.} We compared the integrated {intensities} of IRAM and ALMA CO data that are smoothed to match the FWHM resolution of $12"$. As a result, {We found} no significant difference within {the} error margin, and hence the missing flux {density} is considered to be negligible.

\section{Results}\label{results}
{Figure \ref{fig1}a shows an optical composite image of M33 obtained with the VLT Survey Telescope (VST). {The ALMA FoV includes several H{\sc ii} regions located in a spiral arm near the galactic center.} Figure \ref{fig1}b} shows a large-scale map of ALMA $^{12}$CO($J$ = 2--1) superposed on H$\alpha$ contours. {A giant molecular cloud} (GMC)---M33GMC~37---is mainly located on the western-half of the ALMA FoV with filamentary structures. Some of CO filaments are likely associated with M33SNR~35, which will be described in a {forthcoming} paper (H. Sano et al. in preparation). We also note that a bright H$\alpha$ source with two local peaks {at the positions of ($\alpha_\mathrm{J2000}$, $\delta_\mathrm{J2000}$) $\sim$ ($01^\mathrm{h}33^\mathrm{m}35\fs24$, $30\degree36\arcmin28\farcs0$) and ($01^\mathrm{h}33^\mathrm{m}35\fs20$, $30\degree36\arcmin26\farcs3$)} are superposed on the GMC.

To derive the mass of GMC, we used {the} following equations:
\begin{eqnarray}
M = m_{\mathrm{H}} \mu D^2 \Omega \sum_{i} [N_i(\mathrm{H}_2)],\\
N(\mathrm{H}_2) = X_\mathrm{CO} \cdot W(\mathrm{CO_{10}}),
\end{eqnarray}
where $m_{\mathrm{H}}$ is the mass of hydrogen, $\mu$ is the mean molecular weight of $\sim2.74$, $D$ is the distance to M33 in units of cm, $\Omega$ is the solid angle of each data pixel, $N_i(\mathrm{H}_2$) is the column density of molecular hydrogen for each data pixel $i$ in units of cm$^{-2}$, $X$ is the CO-to-H$_2$ conversion factor in units of (K km s$^{-1}$)$^{-1}$ cm$^{-2}$, and $W$(CO$_\mathrm{10}$) is integrated intensity of $^{12}$CO($J$ = 1--0) line emission. {In the present study, we adopt $D = 2.5 \times 10^{24}$ cm, corresponding to the distance of 817 kpc (\cite{2001ApJ...553...47F})}. Following the previous study by \citet{2012A&A...542A.108G}, we used $X = 4.0 \times 10^{20}$ (K km s$^{-1}$)$^{-1}$ cm$^{-2}$. We also derived a typical intensity ratio of $^{12}$CO($J$ = 2--1) / $^{12}$CO($J$ = 1--0) $\sim0.7$ toward M33GMC~37 through the comparison with an archival $^{12}$CO($J$ = 1--0) cube data obtained with the Nobeyama Radio Observatory 45-m telescope (\cite{2011PASJ...63.1171T}). Finally{, the derived size and mass of the GMCs are} $\sim60$ pc and $\sim5 \times 10^5$ $M_\odot$, respectively. These values are roughly consistent with the previous study by \citet{2012ApJ...761...37M} using the Atacama Submillimeter Telescope Experiment: size of $63 \times 54$ pc and virial mass of $(6.7 \pm 4.8) \times 10^5$ $M_\odot$.

\begin{figure*}
\begin{center}
\includegraphics[width=\linewidth]{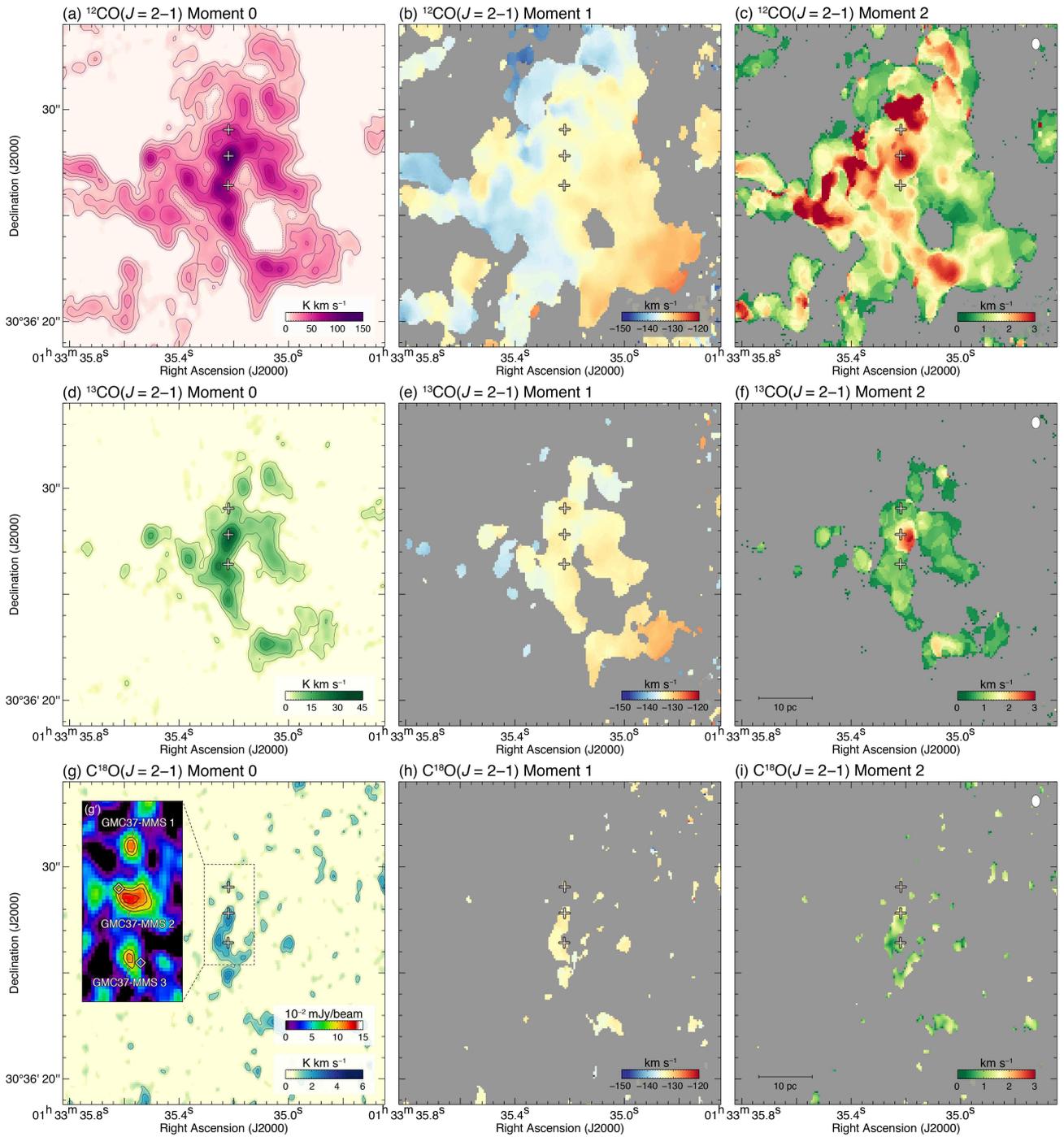}
\end{center}
\caption{{Maps of moment 0 (integrated intensity: Figures \ref{fig2}a, \ref{fig2}d, and \ref{fig2}g), moment 1 (peak velocity: Figures \ref{fig2}b, \ref{fig2}e, and \ref{fig2}h), and moment 2 (velocity dispersion: Figures \ref{fig2}c, \ref{fig2}f, and \ref{fig2}i) toward M33GMC~35 for each line emission. The contour levels of moment 0 maps are 8, 14, 22, 32, 44, 58, 74, and 92 K km s$^{-1}$ for $^{12}$CO($J$ = 2--1); 4, 8, 14, and 22 K km s$^{-1}$ for $^{13}$CO($J$ = 2--1); and 1, 1.5, 2, 3, and 5 K km s$^{-1}$ for C$^{18}$O($J$ = 2--1). Intensity distribution of 1.3 mm continuum is also shown in Figure \ref{fig2}g$'$. The contour levels of {millimeter} continuum are 9.0, 10.5, and 12.0 $\times$ 10$^{-2}$ \red{mJy beam$^{-1}$}. The crosses and diamonds represent the intensity peak positions of {millimeter} continuum sources and H$\alpha$. The beam size and scale bar are also shown in the top right corner and bottom left corner of Figures \ref{fig2}c, \ref{fig2}f, and \ref{fig2}i, respectively.}}
\label{fig2}
\end{figure*}%

{Figures \ref{fig2}a, \ref{fig2}d, and \ref{fig2}g show the maps of moment 0 (integrated intensity) for each line emission toward M33GMC~37. The area shown in Figure \ref{fig2} is indicated by the dashed box in Figure \ref{fig1}.} The $^{12}$CO($J$ = 2--1) clouds are detected as diffuse emission with a ring-like structure in southwest, while the $^{13}$CO($J$ = 2--1) clouds are {only bright in the ring-like structure. The C$^{18}$O($J$ = 2--1)} clouds are concentrated in the central region of GMC where the $^{12}$CO($J$ = 2--1) and {$^{13}$CO($J$ = 2--1) clouds} are bright. {We also find that t}here are three sources in 1.3 mm continuum {as shown in Figure \ref{fig2}g$'$}---{GMC37-}MMSs~1--3 {(hereafter refer to as MMSs~1--3)}---, which are detected at a $5\sigma$ level or higher. The basic physical properties of three continuum sources are listed in Table \ref{tab1}. The peak brightness temperatures of the three sources are comparable, but the spatial extent of MMS~2 is twice as large than that of MMSs~1 and 3. It is possible that these {millimeter} continuum sources are physically related to the H$\alpha$ emission. In particular, two of them (MMSs~2 and 3) are located in the vicinity of the two H$\alpha$ peaks {(as shown by diamonds in Figure \ref{fig2}g$'$)} associated with the brightest CO cloud, indicating that the exciting stars of MMSs~2--3 and H$\alpha$ peaks are the same. These regions are also bright in the $Spitzer$ 24~$\mu$m image (\cite{2007A&A...476.1161V}).

{Figures \ref{fig2}b, \ref{fig2}e, and \ref{fig2}h show the maps of moment 1 (peak velocity). The peak velocities of $^{12}$CO($J$ = 2--1) and $^{13}$CO($J$ = 2--1) clouds are discontinuous with a velocity jump of $\sim$10 km s$^{-1}$ toward the {millimeter} continuum sources and the eastern side of them, while that of C$^{18}$O($J$ = 2--1) line emission shows no significant velocity jump within the clouds. Figures \ref{fig2}c, \ref{fig2}f, and \ref{fig2}i show the maps of moment 2 (velocity dispersion). A larger velocity dispersion above 3 km s$^{-1}$ is seen in several regions of both the $^{12}$CO($J$ = 2--1) and $^{13}$CO($J$ = 2--1) clouds, where the velocity jumps are identified (see also Figure \ref{fig2}b). On the other hand, there is no sign of such larger velocity dispersion in the C$^{18}$O($J$ = 2--1) clouds.}

\begin{figure}
\begin{center}
\includegraphics[width=75mm]{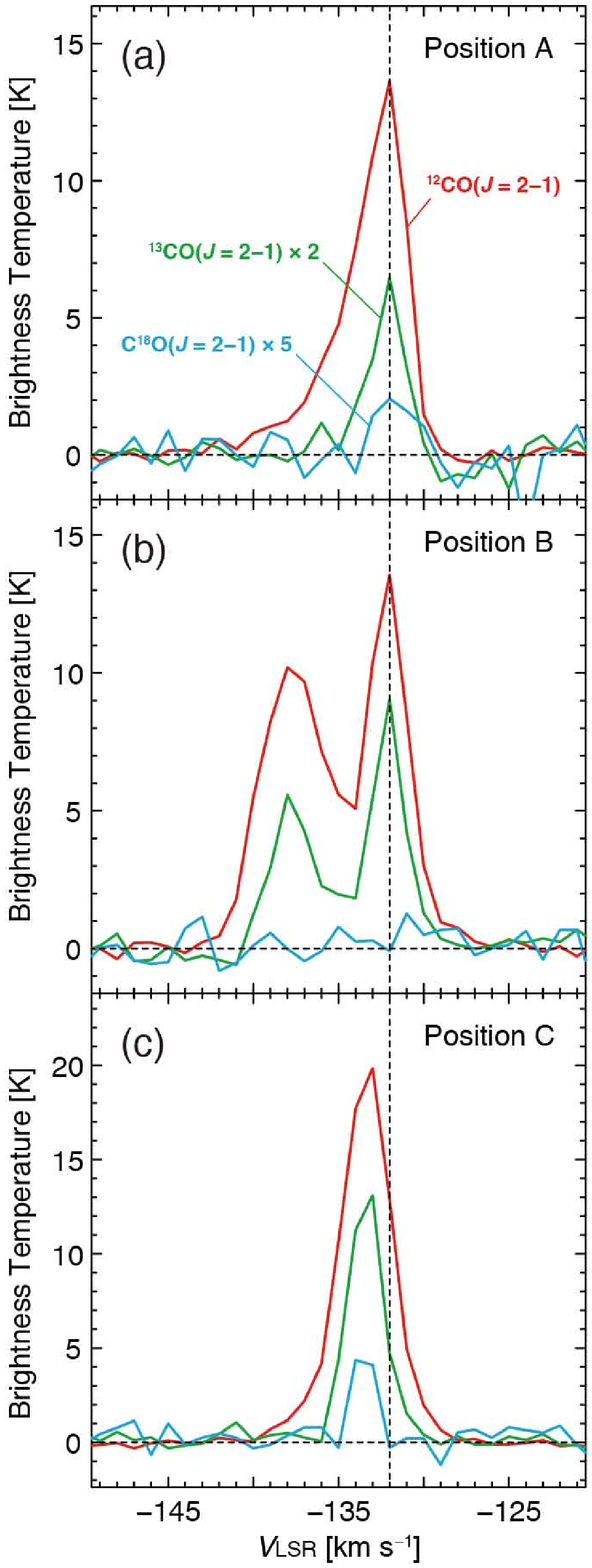}
\end{center}
\caption{{Typical CO spectra in the positions of A [($\alpha_\mathrm{J2000}$, $\delta_\mathrm{J2000}) = (01^\mathrm{h}33^\mathrm{m}35\fs22$, $30\degree36\arcmin29\farcs0$)], B [($\alpha_\mathrm{J2000}$, $\delta_\mathrm{J2000}) = (01^\mathrm{h}33^\mathrm{m}35\fs19$, $30\degree36\arcmin27\farcs6$)], and C [($\alpha_\mathrm{J2000}$, $\delta_\mathrm{J2000}) = (01^\mathrm{h}33^\mathrm{m}35\fs25$, $30\degree36\arcmin26\farcs6$)].} The $^{13}$CO($J$ = 2--1) and C$^{18}$O($J$ = 2--1) brightness temperature are multiplied by a factor of two and five, respectively. {The vertical dashed line indicates the systemic velocity of M33GMC~37.}}
\label{fig0}
\end{figure}%

\begin{table}[t]
\tbl{Physical properties of 1.3 mm continuum sources}{%
\begin{tabular}{lccccc}
\hline 
\hline\noalign{\vskip3pt} 
\multicolumn{1}{c}{Name} & $\alpha_{\mathrm{J2000}}$ & $\delta_{\mathrm{J2000}}$ & {$F_{\rm peak}$} & Size\\
& ($^{\mathrm{h}}$ $^{\mathrm{m}}$ $^{\mathrm{s}}$) & ($^{\circ}$ $\arcmin$ $\arcsec$) & (\red{mJy beam$^{-1}$}) & (pc)\\
\multicolumn{1}{c}{(1)} & (2) & (3) & (4) & (5)\\
\hline\noalign{\vskip3pt} 
{GMC37-MMS~1} & 01 33 35.22 & 30 36 29.0 & 0.12 & 1.7\\ 
{GMC37-MMS~2} & 01 33 35.22 & 30 36 27.8 & 0.13 & 3.0\\ 
{GMC37-MMS~3} & 01 33 35.22 & 30 36 26.4 & 0.11 & 1.6\\ 
\hline\noalign{\vskip3pt} 
\end{tabular}}
\label{tab1}
\begin{tabnote}
\hangindent6pt\noindent
Note. --- Col. (1): Name of {millimeter} continuum source. Cols. (2--3) Positions of the maximum {flux}. Col. (4): {Peak flux of 1.3 mm continuum}. Col. (5): Size of continuum source defined as $(S / \pi)^{0.5} \times 2$, where $S$ is the total surface area of continuum source surrounded by a contour of the $5 \sigma$ level.
\end{tabnote}
\end{table}

{Figure \ref{fig0} shows the typical CO spectra toward three positions A, B, and C. The position~A represents the direction of MMS~1. The positions~B and C correspond the regions with the largest velocity dispersion in $^{13}$CO($J$ = 2--1) line emission (see also Figure \ref{fig2}f) and with the maximum intensity of C$^{18}$O($J$ = 2--1) line emission (see also Figure \ref{fig2}g), respectively. The CO spectra of the position A is centered at $\sim$132 km s$^{-1}$, corresponding to the systemic velocity of the M33GMC~37 region which was derived by a rotation model of the H{\sc i} disk \citep{2010A&A...522A...3G}. We note that the $^{12}$CO($J$ = 2--1) line emission shows a wing-like profile with a velocity extent of more than 5 km s$^{-1}$. For the position B, w}e find double-peak profiles in both the $^{12}$CO($J$ = 2--1) and $^{13}$CO($J$ = 2--1) line emission, suggesting that there are two individual clouds toward the line-of-sight. We hereafter refer to the component at $V_\mathrm{LSR}$ = {$-$}$143.0$--$-134.0$ km s$^{-1}$ as the ``blue cloud'' and that at $V_\mathrm{LSR}$ = $-133.0$--$-127.0$ km s$^{-1}$ as the ``red cloud.'' {On the other hand, the CO spectra of the position~C---the brightest peak in C$^{18}$O($J$ = 2--1)---show a single peak profile for each line emission. Note that t}heir central velocities of $V_\mathrm{LSR} \sim -133.5$ km s$^{-1}$ are roughly corresponding to the mean velocity of red and blue clouds.

\begin{figure*}
\begin{center}
\includegraphics[width=\linewidth]{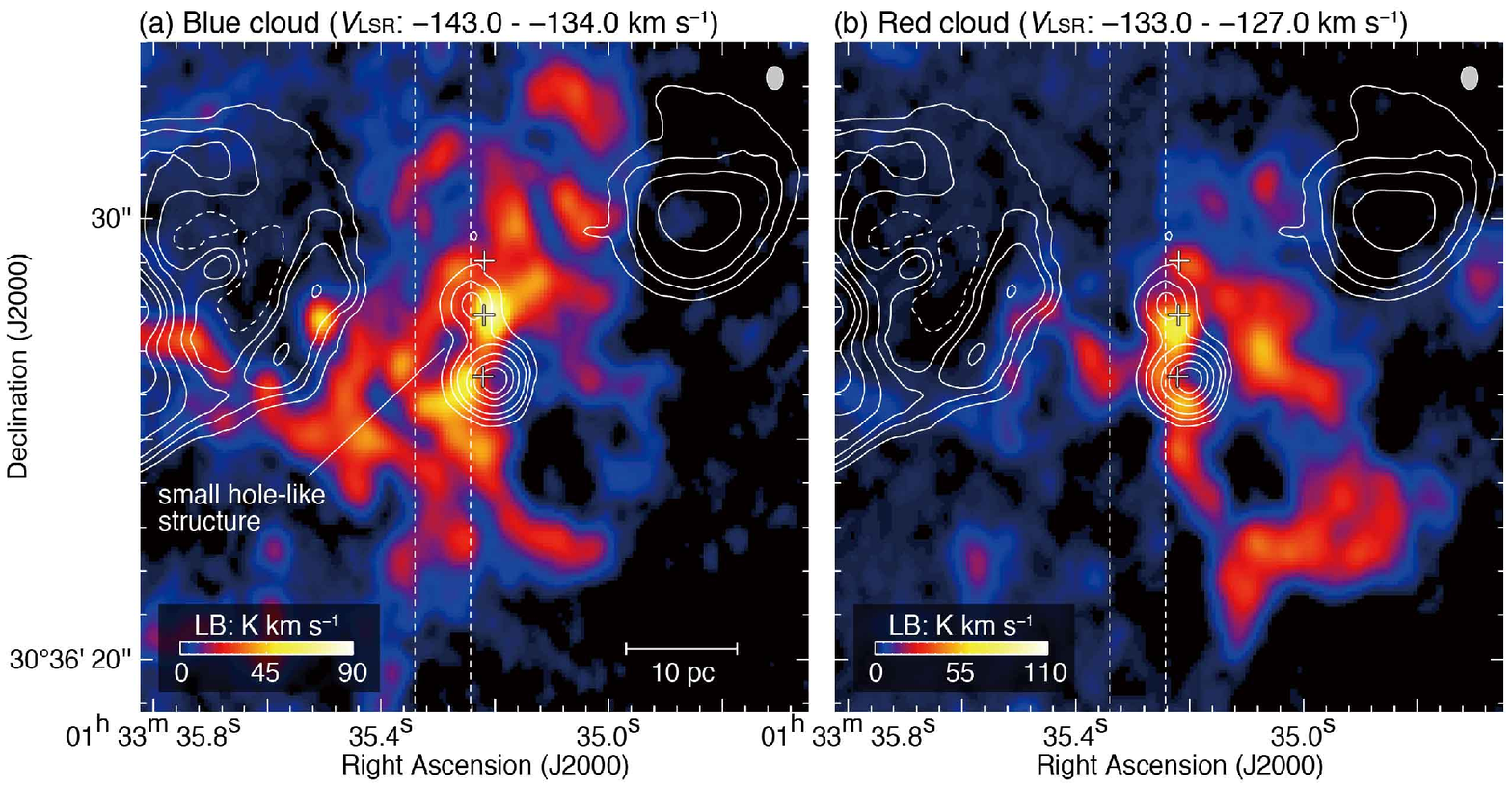}
\end{center}
\caption{Integrated intensity maps of $^{12}$CO($J$ = 2--1). The velocity range is $-143$--$-134$ km s$^{-1}$ for (a) and $-133$--$-127$ km s$^{-1}$ for (b). The vertical dashed lines indicate the integration {range in R.A.} for Figure {\ref{fig4}}. The crosses represent {the} positions of 1.3 mm continuum. {The white contours indicate H${\alpha}$ emission as shown in Figure \ref{fig1}b. The scale bar and beam size are also shown in the bottom and top right corners for each panel.}}
\label{fig3}
\end{figure*}%

Figures \ref{fig3}a and \ref{fig3}b show spatial distributions of blue and red clouds. The blue cloud is elongated in the northwestern direction, while the red cloud is stretched in the southwestern direction. Both clouds have filamentary CO structures. The typical width and length of filaments are $\sim3$ pc and $\sim10$ pc, respectively. The total mass is estimated to be $\sim2.4 \times 10^5$ $M_\odot$ for the blue cloud and $\sim1.9 \times 10^5$ $M_\odot$ for the red cloud using equations (1) and (2). Note that {MMS~2} is likely associated not only with the brightest peak of blue cloud, but also with that of red cloud. {The other continuum sources---MMSs~1 and 3---appear to be overlapped with both the red and blue clouds.} The peak column density of molecular hydrogen is $\sim4.5 \times 10^{22}$ cm$^{-2}$ for the blue-cloud, and $\sim4.1 \times 10^{22}$ cm$^{-2}$ for the red-cloud. We also find a small hole-like structure in the blue cloud centered at ($\alpha_\mathrm{J2000}$, $\delta_\mathrm{J2000}$) $\sim$ ($01^\mathrm{h}35^\mathrm{m}35\fs28$, $30\degree36\arcmin27\farcs1$). Note that the peak positions of H$\alpha$ and {millimeter} continuum are placed on the edge of the hole-like structure, not on the center of {it}.

\begin{figure}
\begin{center}
\includegraphics[width=\linewidth]{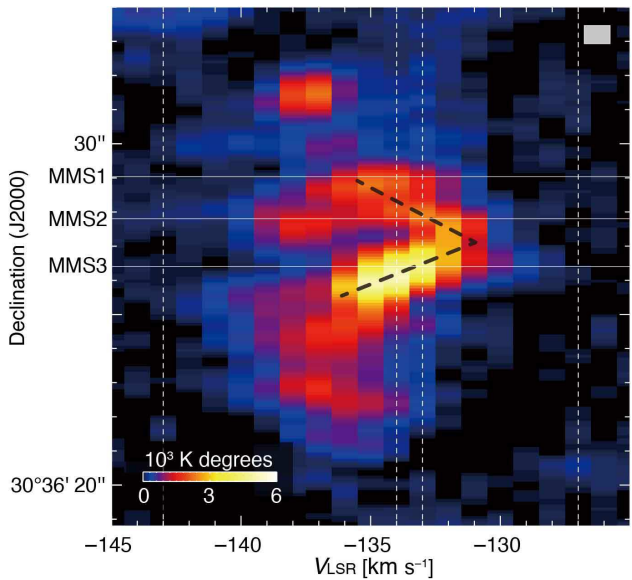}
\end{center}
\caption{Declination--velocity diagram of $^{12}$CO($J$ = 2--1). The integration range is {from $01^\mathrm{h}33^\mathrm{m}35\fs243$ to $01^\mathrm{h}33^\mathrm{m}35\fs341$}. The vertical lines indicate integration velocity ranges for Figure \ref{fig3}. {The three {horizontal solid lines} indicate the Declination positions of MMS{s}~1--3.} The beam size is shown in the top right corner.}
\label{fig4}
\end{figure}

Figure \ref{fig4} shows the position--velocity diagram of $^{12}$CO($J$ = 2--1). {The integration range in Right Ascension is from { $01^\mathrm{h}33^\mathrm{m}35\fs243$ to $01^\mathrm{h}33^\mathrm{m}35\fs341$}, corresponding to the spatial extent of the small hole-like structure appeared in the blue cloud (see also Figure \ref{fig3}a). The blue cloud is clearly seen whose radial velocity is centered at $V_\mathrm{LSR} \sim -137$ km s$^{-1}$. We find a V-shaped structure in the position--velocity diagram (see dashed black lines in Figure \ref{fig4}). The center velocity of red cloud ($V_\mathrm{LSR} \sim -132$ km s$^{-1}$) is corresponding to the vertex of the V-shaped structure. The cavity-like structure and intensity peak of $^{12}$CO($J$ = 2--1) are seen at ($\delta_\mathrm{J2000}$, $V_\mathrm{LSR}$) $\sim$ ($30\degree36\arcmin27\farcs1$, $-135.0$ km s$^{-1}$) and $\sim$ ($30\degree36\arcmin26\farcs0$, $-134.0$ km s$^{-1}$), respectively.}

\section{Discussion}
\subsection{{Presence of high-mass stars and their spectral types}}\label{spectral}

\begin{figure*}
\begin{center}
\includegraphics[width=\linewidth]{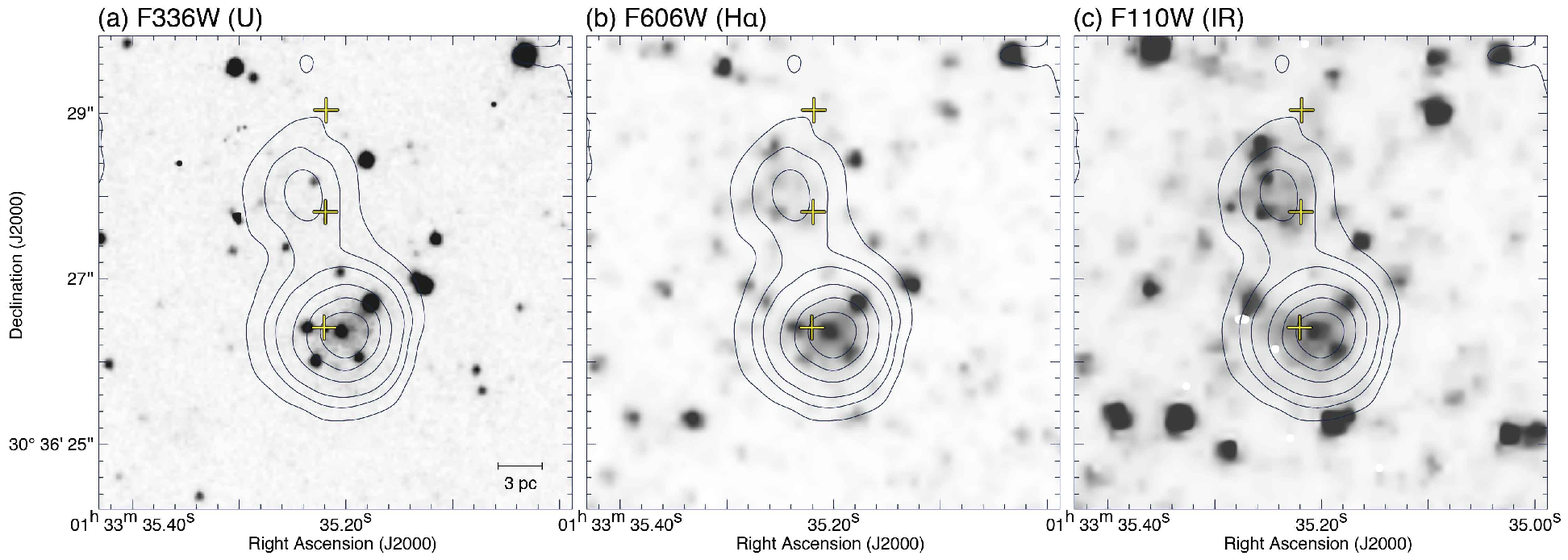}
\end{center}
\caption{{Maps of optical images obtained with the {\it Hubble Space Telescope} ({\it HST}): (a) {\it HST}/WFC3 map with F336W filter (U-band), (b) {\it HST}/WFPC2 map with F606W filter (H$\alpha$), and (c) {\it HST}/WFC3 map with F110W filter (IR-band). The superposed contours and crosses are the same as in Figure \ref{fig3}. The scale bar is also shown in the bottom right corner in (a).}}
\label{fig5}
\end{figure*}

{Although there is no previous study of the stars embedded within the H{\sc ii} region in M33,} detection of bright H$\alpha$ emission, 1.3~mm continuum sources MMSs~1--3, and 24~$\mu$m emission indicates {several} {high-mass} stars are {associated with} the {two} molecular cloud{s}. {To confirm the presence of {high-mass} stars, we compared them with archival optical datasets obtained with the {\it Hubble Space Telescope} ({\it HST}). Figure \ref{fig5} shows the high-spatial resolution optical images obtained with the {\it HST} WFPC2 and WFC3 detectors. {The coordinate offsets of all datasets obtained with {\it{HST}} were corrected by comparing the position of the bright point sources with those catalogued in {\it{Gaia}} Data Release 2 \citep{2018A&A...616A...1G}.} We can clearly see $\sim10$ {high-mass} stars bright in the U-band, H$\alpha$, and IR-band within the {Kitt Peak National Observatory (KPNO)} H$\alpha$ contours.  Hereafter, we assume that ten {high-mass} stars are associated with the H$\alpha$, 1.3 mm continuum sources, and 24~$\mu$m emission.}

According to \citet{2007A&A...476.1161V}, the luminosity of {the} 24~$\mu$m emission is $\sim1.7 \times 10^{38}$ erg s$^{-1}$ and that of {the} H$\alpha$ emission is $\sim4.6 \times 10^{37}$ erg s$^{-1}$ using the $Spitzer$ MIPS and KPNO H$\alpha$ datasets. To estimate the spectral types of {high-mass} stars, we used two different methods. One {estimates the} total infrared luminosity $L$(TIR) using the {measured} 24~$\mu$m luminosity $L$(24~$\mu$m) and {the} following equation (\cite{2007A&A...476.1161V}):  
\begin{eqnarray}
\log L(\mathrm{TIR}) = \log L(\mathrm{24\;} \mu \mathrm{m}) + 0.908
\label{eq3}
\end{eqnarray}
The total infrared luminosity of {the} {high-mass} stars is estimated to be $\sim1.4 \times 10^{39}$ erg s$^{-1}${, which} corresponds to ten {O7.5V} stars (\cite{2005A&A...436.1049M}), assuming that the {ten} {high-mass} stars have the same spectral types.

The other {method} is utilized Lyman continuum luminosities $N_\mathrm{Lyman}$ (in units of photons) that are derived from extinction corrected H$\alpha$ luminosities $L_{\mathrm{H}\alpha}$ (in units of erg s$^{-1}$) using {the} following equation (\cite{1996AJ....111.1252M}):
\begin{eqnarray}
N_\mathrm{Lyman} = 7.3 \times 10^{11} L_{\mathrm{H}\alpha}
\label{eq4}
\end{eqnarray}
We then obtain $N_\mathrm{Lyman}$ = $1.1 \times 10^{49}$ photons, corresponding to {ten B0V} stars (\cite{2005A&A...436.1049M}). To summaries, the spectral types of {high-mass} stars embedded within M33GMC~37 are estimated to be {B0V--O7.5V} assuming that there are {ten} {high-mass} stars with the same spectral types. Further detailed photometric and spectroscopic observations are needed to clarify {the} number of {high-mass} stars and their spectral types.

\subsection{{Can stellar feedback explain the large velocity separation of the two clouds?}}\label{feedback}
In the present study, we spatially resolved filamentary CO structures of M33GMC~37 using ALMA with spatial resolution of $\sim2$ pc ($\Delta \theta \sim 0.5"$). There are two individual molecular clouds---red and blue clouds---with a velocity separation of $\sim6$ km s$^{-1}$ and the mass of $\sim2 \times 10^5$ $M_\odot$ for each. The densest part of the GMC is significantly detected in C$^{18}$O($J$ = 2--1), containing up to $\sim10$ high-mass stars with the spectral types of B0V--O7.5V. {In this section, we shall discuss whether the velocity separation of two clouds can be explained by acceleration due to the stellar feedback.}

{First, we claim that the total momentum of stellar winds is significantly lower than that of the two molecular clouds. The typical wind momentum of an O-type star is calculated to be $\sim$$2 \times 10^{3}$ $M_\odot$ km s$^{-1}$, by considering a mass accretion rate of $\sim$$10^{-6}$ $M_\odot$ yr$^{-1}$, a wind duration time of $\sim$10$^6$ yr, and a wind velocity of $\sim$2,000 km s$^{-1}$ (see \cite{2000ARA&A..38..613K} and references therein). In the case of M33GMC~37, the total wind momentum is therefore to be $\sim$$2 \times 10^{4}$ $M_\odot$ km s$^{-1}$ assuming that there are ten O-type stars (see Section \ref{spectral}). On the other hand, we derive a total cloud momentum of $\sim$$1.3 \times 10^6$ $M_\odot$ km s$^{-1}$. Here we assume that the two clouds show an expanding motion with the expanding velocity of $\sim$3 km s$^{-1}$, corresponding to a half value of the velocity separation of the two clouds. Since the total cloud momentum is roughly two orders magnitude higher than the total wind momentum, we conclude that the velocity separation of two clouds cannot be explained by the expanding gas motion due to the stellar feedback.}

\begin{figure*}
\begin{center}
\includegraphics[width=\linewidth]{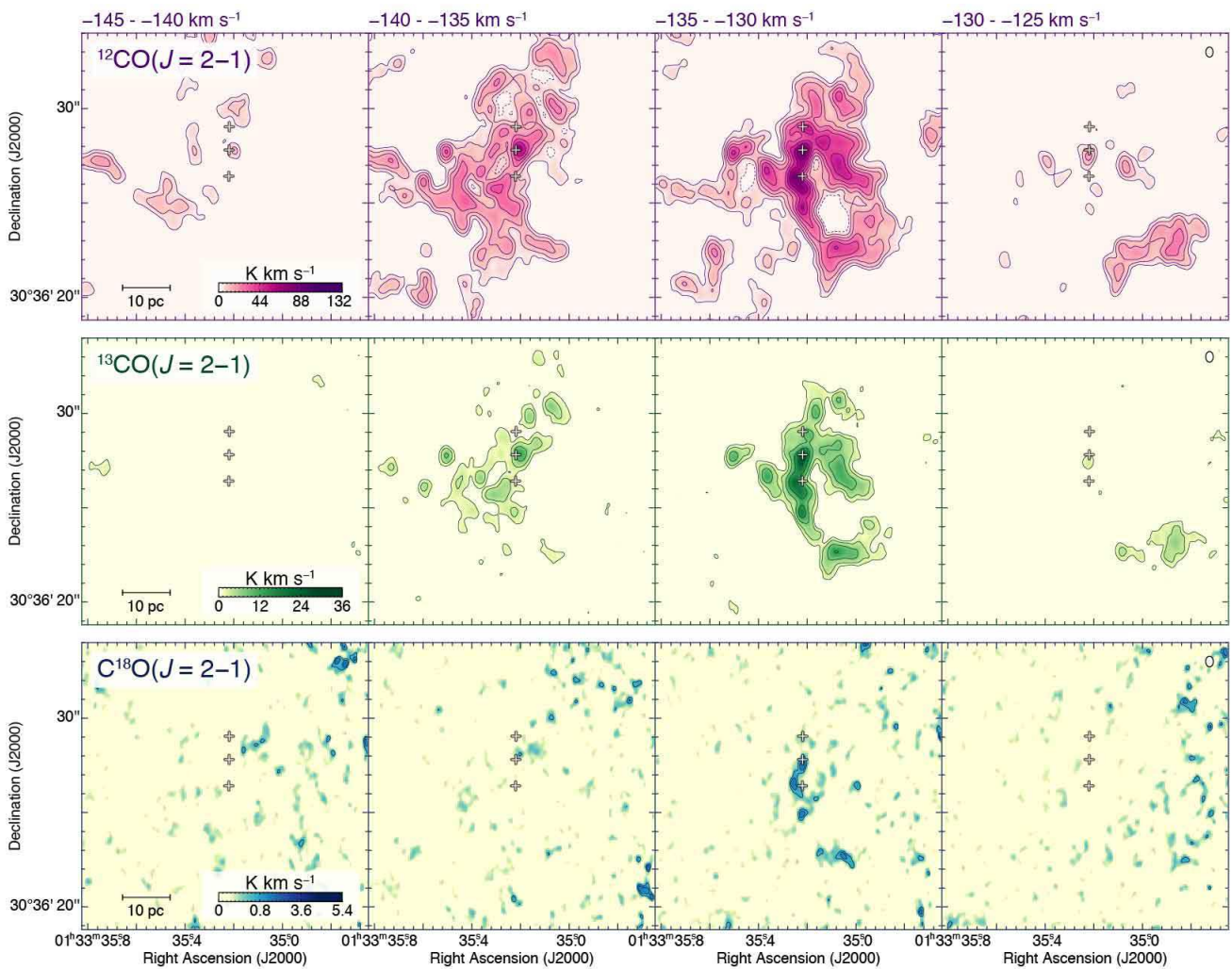}
\end{center}
\caption{{Velocity channel maps of (a) $^{12}$CO($J$ = 2--1), (b) $^{13}$CO($J$ = 2--1), and C$^{18}$O($J$ = 2--1) line emission toward M33GMC~37. Each panel shows intensity distribution integrated every 5 km s$^{-1}$ in a velocity range from $-145$ to $-125$ km s$^{-1}$. The contour levels are 4, 8, 14, 22, 32, 44, 58, and 74 K km s$^{-1}$ for $^{12}$CO($J$ = 2--1); 2, 4, 8, and 14 K km s$^{-1}$ for $^{13}$CO($J$ = 2--1); and 1.5, 2, 3, and 5 K km s$^{-1}$ for C$^{18}$O($J$ = 2--1). The beam size and scale bar are also shown in the top right corner and bottom left corner, respectively. The crosses indicate the positions of the {millimeter} continuum sources.}}
\label{figa}
\end{figure*}

{This interpretation is also consistent with the velocity distributions of the two clouds. If the expansion of clouds is centered at an exciting star, a ring-like gas distribution will be expected not only in the spatial distribution such as the moment 0 map, but also in the position--velocity diagram (c.f., Figure 8 in \cite{2015ApJ...806....7T}). For M33GMC~37, however, we could not find such distributions around the three {millimeter} continuum sources (see Figures \ref{fig0}a and \ref{fig4}). We also have created velocity channel maps in order to justify the velocity and spatial distributions of clouds, because expanding clouds will be spatially shifted from inward to outward in the velocity channel maps (e.g., \cite{2012ApJ...746...82F,2017JHEAp..15....1S}). Figure \ref{figa} shows the velocity channel maps for each CO line emission. We confirmed that there are no clear signs of such spatial shifts with increasing velocity. On the other hand, we note that a tiny CO clump near MMS~2 appears in all velocity ranges of $^{12}$CO($J$ = 2--1) line emission, suggesting a possible evidence for the outflowing gas as well as the wing-like profile of MMS~1 (Figure \ref{fig2}a). Although the candidate of outflowing gas is very concentrated within 1~pc scales, the outflowing gas could not strongly affect to the dynamical motion of the two clouds. To summarize, the stellar feedback effect is negligible for the cloud kinematics, and therefore the velocity separation of two clouds cannot be explained by acceleration by stellar winds and outflowing gas.}

\subsection{{An Alternative Idea:} High-mass Star Formation Triggered by Cloud-Cloud Collisions in M33GMC~37}\label{ccc}
{In the previous section, we conclude that the velocity separation of two clouds have not caused by the stellar feedback effect. It is therefore reasonable to interpret that the individual two clouds have collided with each other and have formed the high-mass stars embedded within M33GMC~37. In this section, we discuss this second scenario.}

To form {high-mass} stars via cloud-cloud collisions, a supersonic velocity separation of two colliding clouds is essential. According to magnetohydrodynamical numerical simulations, the effective {J}eans mass in the shock-compressed layer is proportional to the third power of {the} effective sound speed {(\cite{2013ApJ...774L..31I}). Here, the effective sound speed is defined as $< c_\mathrm{s}^2 + c_\mathrm{A}^2 + \Delta v^2 >^{0.5}$, where $c_\mathrm{s}$ is the sound speed, $c_\mathrm{A}$ is the Alfven speed, and $\Delta v$ is the velocity dispersion.} A supersonic velocity separation {of} at least a few km s$^{-1}$ therefore produces a large mass accretion rate on the order of $\sim10^{-4}$--$10^{-3}$ $M_\odot$ yr$^{-1}$, which allows mass growth of stars against the stellar feedback (\cite{2013ApJ...774L..31I}). For the case of M33GMC~37, the observed velocity separation of red and blue clouds is $\sim6$ km s$^{-1}$ (see CO spectra in Figure \ref{fig2}b). {Although the velocity separation will be changed due to the projection effect,} this is roughly consistent with the typical velocity separation in the Galactic {high-mass} star forming regions triggered by cloud-cloud collisions: e.g., M20 ($\sim7.5$ km s$^{-1}$, \cite{2011ApJ...738...46T}), RCW 36 ($\sim5$ km s$^{-1}$, \cite{2018PASJ...70S..43S}), M42 ($\sim7$ km s$^{-1}$, \cite{2018ApJ...859..166F}), and RCW~166 ($\sim5$ km s$^{-1}$, \cite{2018PASJ...70S..47O}).

We next focus on the velocity structures of colliding two clouds. Previous numerical simulations and observational results demonstrated that colliding clouds create an intermediate velocity component---bridging feature---connecting two clouds due to the deceleration by collisions, if the projected velocity separation is significantly larger than the linewidth of colliding clouds (see \cite{2018ApJ...859..166F} and references therein). When the two colliding clouds have roughly {the} same column density of gas, the two clouds will be merged and {appears as} a single peak CO profile centered at the mean velocity of {the} two colliding clouds (e.g., \cite{2017PASJ...69L...5F}). In the case of M33GMC~37, {the $V_\mathrm{LSR} \sim -135$ km s$^{-1}$ feature in the $^{13}$CO spectrum at the position A is possibly a bridging feature (see Figure \ref{fig2}b), but is not significantly detected because of {the} small velocity separation of {the} two colliding clouds.} {However}, CO spectra in the position B---{the} densest part of the {colliding} clouds---show the single peak CO profiles centered at $V_\mathrm{LSR} \sim -133.5$ km s$^{-1}$, roughly corresponding to the mean velocity of {the} red and blue clouds (see also Figure \ref{fig2}c). It is consistent that the two clouds have roughly {the} same column density of $\sim4$--$5 \times 10^{22}$ cm$^{-2}$, assuming the cloud-cloud collision {has} occurred.

\begin{figure*}
\begin{center}
\includegraphics[width=\linewidth]{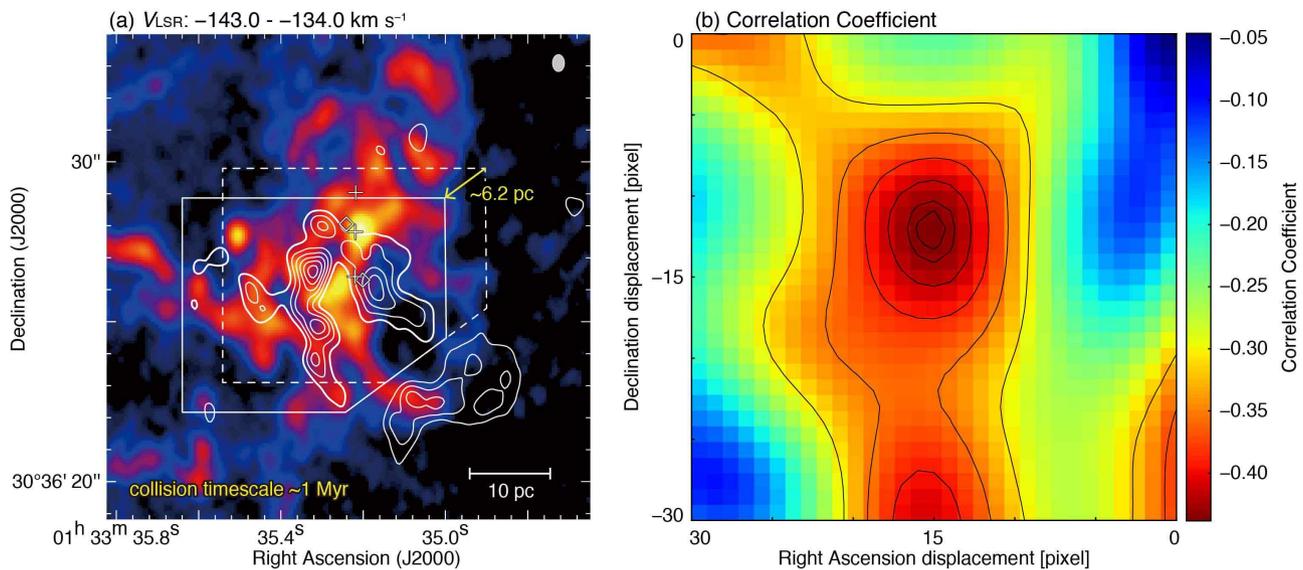}
\end{center}
\caption{(a) Complementary spatial distribution of red cloud (contours) and blue clouds (image). The velocity ranges of image and contours are the same as shown in Figures \ref{fig3}a and \ref{fig3}b, respectively. The contour levels are 16, 26, 36, 46, 56, and 66 K km s$^{-1}$. The contours are spatially displaced {$\sim$6.2} pc in the direction of southeast {(P. A. = 137 degree)}. The dashed and solid rectangles indicate before and after displacement of the contours, respectively. The crosses and diamonds are the same as shown in Figure \ref{fig2}. The scale bar and beam size are also shown in the bottom and top right corners, respectively. (b) Map of Pearson's correlation coefficient as a function of displacements in Right Accession direction and in the Declination direction in the unit of pixel. The size of a pixel is 84.5 mas. \red{The origin of coordinates is ($\alpha_\mathrm{J2000}$, $\delta_\mathrm{J2000}) = (01^\mathrm{h}33^\mathrm{m}35\fs22$, $30\degree36\arcmin27\farcs5$).} The intensity distributions of $^{12}$CO($J$ = 2--1) enclosed by the dashed polygon (blue cloud) and the solid polygon (red cloud) were used to calculate the correlation coefficient. \red{Superposed contours indicate the values of  correlation coefficients, whose levels are from $-0.435$, $-0.430$, $-0.415$, $-0.390$, $-0.355$, and $-0.310$.}}
\label{fig6}
\end{figure*}%

The V-shaped structure in the position--velocity diagram as shown in Figure \ref{fig4} provides us with suggestive evidence for \red{head-on} cloud-cloud collision. As discussed above, an intensity depression or a hole-like structure might indicate the spot of collision between two clouds \red{for the case of the head-on collision}. If we make a position--velocity diagram that includes the collision region, we can find not only the bridging feature but also the V-shaped structure in the position--velocity diagram (e.g., \cite{2018ApJ...859..166F}, \yearcite{2018PASJ...70S..44F}, \yearcite{2018PASJ...70S..46F}; \cite{2018PASJ...70S..48H,2018PASJ..tmp..121T}, \yearcite{2018PASJ...70S..51T}; \cite{2019PASJ..tmp...46F}). \red{Note that if the oblique collision---e.g., collisions with the edges of the clouds---occurred, we could not find such a V-shaped structure in the position--velocity diagram \citep{2020PASJ..tmp..163F}.} For M33GMC 37, we can clearly see the {V-shaped structure connecting the red and blue clouds as the intermediate velocity component {that} is called the bridging feature. Moreover, {the} presence of {an} intensity peak and depression in the V-shaped structure is predicted by the synthetic observations of a theoretical result of \red{the head-on} cloud-cloud collision (e.g., \cite{2018PASJ...70S..44F}).}

{Another important} signature of a collision is complementary spatial distribution of colliding clouds. In general, colliding clouds are not the same size, such as simulated by \citet{1992PASJ...44..203H} and \citet{2010MNRAS.405.1431A}. In fact, observational results indicate that colliding clouds have different {sizes, morphologies, and density distributions} (e.g., \cite{1994ApJ...429L..77H,2009ApJ...696L.115F,2014ApJ...780...36F}, \yearcite{2018ApJ...859..166F}, \yearcite{2018PASJ...70S..41F}; \cite{2018PASJ...70S..49E}, {\cite{2018ApJ...862....8T}}, \cite{2019ApJ...875..138D}, \yearcite{2019ApJ...878...26D}). In such {cases}, one of the colliding clouds can create a hole-like structure in the other cloud, if the colliding cloud has a denser part and/or smaller size than the other cloud. This produces the complementary spatial distribution of two clouds with different systematic velocity. Furthermore, the angle of two colliding clouds $\theta$ is generally not 0 degree{s} or 90 degree{s} relative to the line-of-sight. It means that we can also observe a spatial displacement between the complementary distributions of two clouds.

To evaluate such complimentary distributions of colliding two clouds, it is useful to calculate the correlation coefficient between the intensity distributions of two clouds (c.f., S. Fujita et al. in preparation). If the correlation coefficient between the two colliding clouds shows a negative value, the two clouds show complementary spatial distributions. The lower the correlation coefficient, the higher is complementarity. By changing values of spatial displacement for one of the colliding clouds, we can obtain a map of correlation coefficient between the two clouds as a function of displacement. The best-fit value of the displacement for each coordinate can be derived as the point of the local minimum.

In M33GMC~37, we find complementary spatial distributions of red and blue clouds {by following the above method}. Figure {\ref{fig6}a} shows the map of blue cloud superposed on the red cloud contours with the spatial displacement of {$\sim6.2$} pc toward the direction of southeast. The brightest peak of {the} red cloud is {fits within} the hole-like structure of {the} blue cloud. {The second and third} minor peaks of {the red} cloud also show good {complementary} distributions {(or anti-correlation)} {with the blue cloud}. {Figure \ref{fig6}b shows the map of correlation coefficient between the two clouds as the function of spatial displacement of the red cloud in the directions of Right Accession and Declination. \red{Since the cloud-collision in M33GMC~37 is likely the head-on collision, the correlation coefficients were calculated toward inner parts of the blue cloud.} We can find the local minimum with the correlation coefficient of $-$0.44 at a shift by 15 pix in R.A. and $-12$ pix in Dec., which is significantly lower than the correlation coefficient of $-0.05$ without displacement. \red{We also caliculate P-value which provides information about whether a statistical hypothesis is significant or not. As a result, the correlation coefficient of $-$0.44 is significant at the 0.5\% significance level.}}

{If the cloud-collision scenario is correct, w}e {can} also estimate the collision time scale {as a dynamical time scale using the values of displacement and velocity difference of two clouds} to be {(spatial displacement) / (velocity difference) = 8.8 pc / 8.5 km s$^{-1}$ $\sim1$} Myr assuming the collision angle $\theta$ = 45 degree. {Since GMCs are thought to be dispersed within $\sim$1.5~Myr due to strong stellar winds and UV radiation \citep{2019Natur.569..519K}, the shorter collision time scale} is consistent with the presence of dense molecular clouds surrounding the {high-mass} stars and small H{\sc ii} regions with a few pc extent in H$\alpha$ emission because of young {age} of the high-mass stars (see Figures \ref{fig1}{b and \ref{fig3}}).

We also note that the number of {high-mass} stars ($\lesssim 10$) in M33GMC~37 is consistent with previous observational studies of cloud-cloud collisions. According to \citet{2018ApJ...859..166F}, the formation of super star clusters containing more than {ten} O-type stars requires collisions of two dense clouds, one of which has a high column density {of} at least $\sim10^{23}$ cm$^{-2}$ (e.g., RCW~38, \cite{2016ApJ...820...26F}; NGC~3603, \cite{2014ApJ...780...36F}; Westerlund~2, \cite{2009ApJ...696L.115F}). On the other hand, the formation of single or a few O-type stars happen{s} in a collision between molecular clouds with low column density of {several} $10^{22}$ cm$^{-2}$ or less (e.g., RCW~120, \cite{2015ApJ...806....7T}; NGC~2359, \cite{2017arXiv170808149S}; {S44, \cite{2018PASJ..tmp..126K}; N4, \cite{2019ApJ...872...49F}}, more detailed results are summarized in {\cite{2019PASJ..tmp..127E}}). For M33GMC 37, the two colliding clouds have a low column density of $\sim$4--$5 \times 10^{22}$ cm$^{-2}$. Therefore, cloud-cloud collisions in M33GMC~37 can create $\sim$10 high-mass stars at most. This is consistent with $\sim$10 stars that are detected by {\it HST} optical images within the KPNO H$\alpha$ boundary (see Figure \ref{fig5}).

\begin{figure}
\begin{center}
\includegraphics[width=\linewidth]{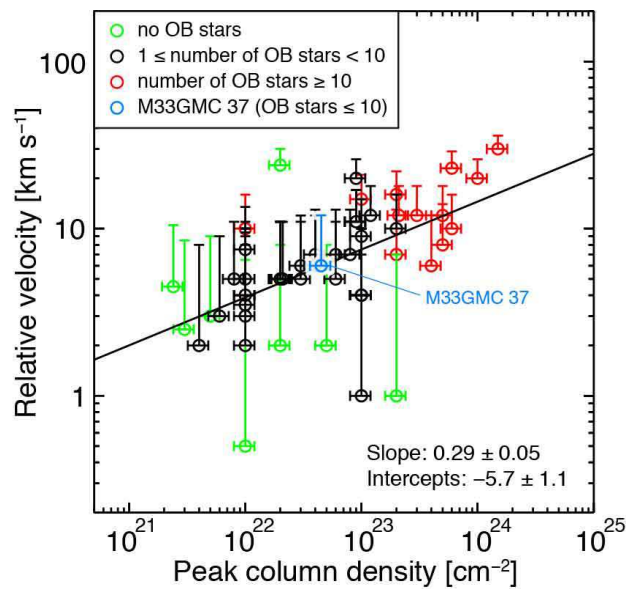}
\end{center}
\caption{{Scatter plot between the peak column density of molecular hydrogen and relative velocity of colliding two clouds toward the sites of cloud-cloud collisions in the Milky Way \citep{2019PASJ..tmp..127E}. The green, black, and red circles represent the H{\sc ii} regions associated with no OB stars, less than ten OB stars, and more than ten OB stars, respectively. The solid line indicates the regression line obtained by least-squares fitting \citep{2019PASJ..tmp..127E}. We also added the data point of M33GMC~37 as the cyan circle.}}
\label{fig8}
\end{figure}%

{In addition, \citet{2019PASJ..tmp..127E} recently revealed that the number of OB stars formed in cloud collisions depends not only on the column density, but also on the velocity difference of the two colliding clouds. Figure \ref{fig8} shows the correlation plot between the column density and relative velocity toward the sites of cloud-cloud collisions in the Milky Way \citep{2019PASJ..tmp..127E}. The authors found that too fast or slow cloud collisions cannot form OB stars even if the cloud column density is high enough. In the case of M33GMC~37, the observational parameters are consistent with the trend seen in previous studies. This possibly indicates that there is no critical difference between the Milky Way and M33 from the view point of high-mass star formation triggered by cloud-cloud collisions. To evaluate the interpretation, more evidence on triggered high-mass star formation via cloud-cloud collisions is needed. Future  ALMA observations of young massive clusters in M33 will allow us to study detailed processes of high-mass star formation.}

\begin{figure*}
\begin{center}
\includegraphics[width=\linewidth]{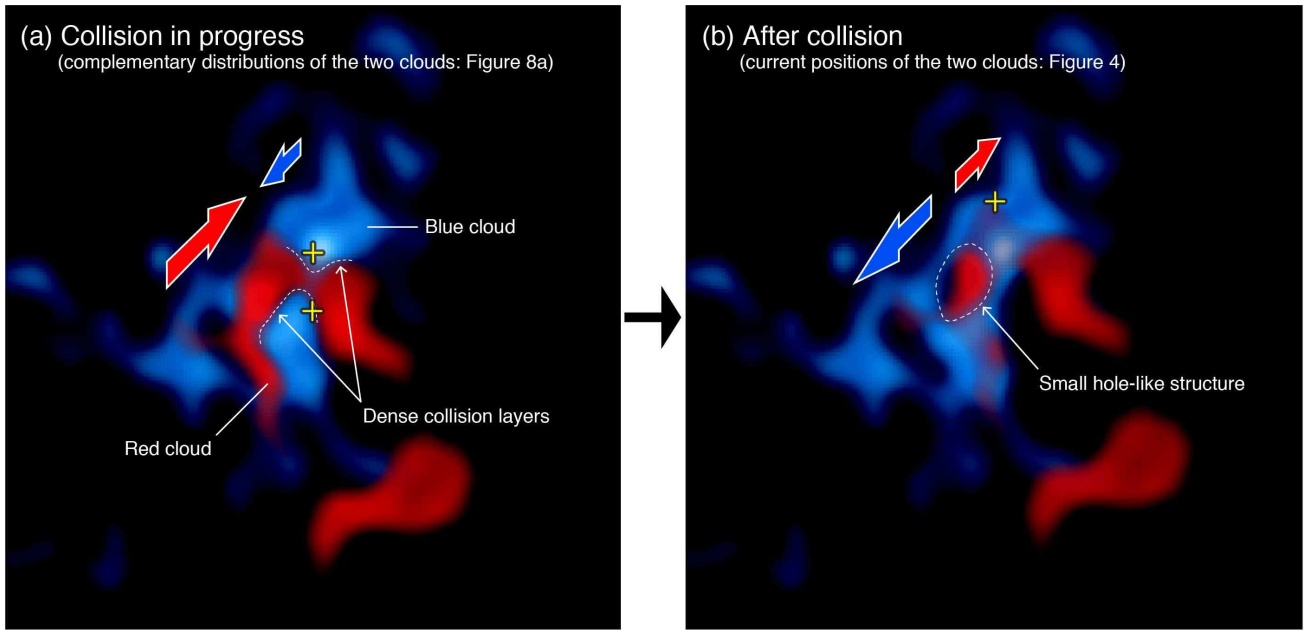}
\end{center}
\caption{{Schematic images of the cloud-cloud collisions in M33GMC~37 for the phases of (a) collision in progress and (b) after collision. The red and blue colors represent the red and blue clouds, respectively. The red and blue arrows indicate the moving directions of the red and blue clouds, respectively. The dense collision layers and the small hole-like structure in the blue cloud are also indicated. The yellow crosses indicate the positions of MMSs~1--3.}}
\label{fig10}
\end{figure*}%

{Finally, we present a summary picture of the high-mass star formation triggered by the cloud collisions in M33GMC~37. Figure \ref{fig10} shows schematic images of the cloud-cloud collisions in M33GMC~37. First, the red and blue clouds were moving in the direction approaching each other with a relative velocity of more than 6 km s$^{-1}$ respect to the line of sight. Next, the two clouds collided with each other $\sim$1 Myr ago. The continuous collisions created the complementary spatial distributions between the two clouds (Figures \ref{fig10}a and \ref{fig6}a). The molecular gas was efficiently compressed especially in the dense collision layers, which formed the high-mass stars toward MMSs~2 and 3 (yellow crosses). Finally, the red cloud penetrated the blue cloud making the small hole-like structure and moved behind the blue cloud, which corresponds to the present positions of the two clouds (Figures \ref{fig10}b and \ref{fig3}). The high-mass star in MMS~1 might have formed recently when the edge of the red cloud finally reached the position of MMS~1. The bright spot and cavity structure in the position--velocity diagram are also consistent with the positions of the dense collision layers and the hole-like structure in the blue cloud, respectively (Figure \ref{fig4}).}

{To summarize}, {the} red and blue clouds in M33GMC~37 fulfill four requirements of {high-mass} {star} formation triggered by cloud-cloud collisions as follows: (1) a super sonic velocity separation of two clouds, (2) complementary spatial distribution with a displacement of colliding clouds, (3) a presence of {a} bridging feature connecting {the} two clouds in velocity space, and (4) V-shaped structure in the position--velocity diagram. We therefore {suggest} that the {high-mass} stars corresponding to ten B0V--O7.5V {types} in M33GMC~37 were formed by cloud-cloud collisions. Further ALMA observations of M33 GMCs will allow us to study {high-mass} star formation via the cloud-cloud collision{s} in the spiral galaxy, the results of which can be directly compared with that of the Milky Way {and Magellanic Clouds}.

\section{Conclusion}\label{conclusion}
In the present study, we carried out new CO($J$ = 2--1) and continuum observations of M33GMC~37 using ALMA with the angular resolution of $\sim0\farcs5$, corresponding to the spatial resolution of $\sim2$ pc at the distance of M33. We revealed two individual molecular clouds with a velocity separation of $\sim6$ km s$^{-1}$ {that} are associated with {up to $\sim10$} {high-mass} stars having the spectral types of {B0V--O7.5V}. The two molecular clouds show complementary spatial distribution with the spatial displacement of {$\sim6.2$} pc. The intermediate velocity component of {the} two clouds {as the} V-shaped structure in the position--velocity diagram {is} also detected. We {propose} a possible scenario that the {high-mass} stars in M33GMC~37 were formed by cloud-cloud collisions approximately {1} Myr ago.

\begin{ack}
This paper makes use of the following ALMA data: ADS/JAO. ALMA\#2018.1.00378.S. ALMA is a partnership of ESO (representing its member states), NSF (USA) and NINS (Japan), together with NRC (Canada) and {NSC} and ASIAA (Taiwan) and KASI (Republic of Korea), in cooperation with the Republic of Chile. The Joint ALMA Observatory is operated by ESO, AUI/NRAO, and NAOJ. {Based on observations made with the NASA/ESA Hubble Space Telescope, and obtained from the Hubble Legacy Archive, which is a collaboration between the Space Telescope Science Institute (STScI/NASA), the Space Telescope European Coordinating Facility (ST-ECF/ESA) and the Canadian Astronomy Data Centre (CADC/NRC/CSA).} This study was financially supported by Grants-in-Aid for Scientific Research (KAKENHI) of the Japanese Society for the Promotion of Science (JSPS, grant Nos. 16K17664, 18J01417, and 19K14758). K. Tokuda was supported by NAOJ ALMA Scientific Research Grant Number of 2016-03B. \red{H.S. was also supported by the ALMA Japan Research Grant of NAOJ Chile Observatory (grant No. NAOJ-ALMA-226).} M.S.\ acknowledges support by the Deutsche Forschungsgemeinschaft through the Heisenberg professor grants SA 2131/5-1 and 12-1. We are also grateful to the anonymous referee for useful comments which helped the authors to improve the paper.
\end{ack}

\end{document}